# Increasing efficiency of a wireless energy transfer system by spatial translational transformation

Shichao Li[1], Fei Sun[1, *], Di An[1], and Sailing He[1, 2, *]

1 Centre for Optical and Electromagnetic Research, Zhejiang Provincial Key Laboratory for Sensing Technologies, JORCEP, East Building #5, Zhejiang University, Hangzhou 310058, China
2 Department of Electromagnetic Engineering, School of Electrical Engineering, Royal Institute of Technology (KTH), S-100 44 Stockholm, Sweden
*corresponding authors: sunfei@zju.edu.cn or sailing@kth.se

**Abstract**
An optical translational projector (OTP) designed by transformation optics is applied to improve the energy transfer efficiency in a wireless energy transfer (WET) system. Numerical simulation results show our OTP can greatly enhance the energy transfer efficiency (e.g. nearly 2 orders, compared to the case without our OTP) in WET systems, which is much larger than previous methods (e.g. magnetic super-lens). A 3D reduced OTP composed by layered isotropic magnetic materials is designed, whose performance has been verified by 3D numerical simulations in 10*MHz*. We also study the influence of loss of metamaterials on the performance of proposed OTP.

## Introduction

Effort to transfer energy wirelessly was made in an early human history, which begins from Nikola Tesla [1]. Wireless energy transmission technology has many important applications, e.g. charging biomedical implants [2], unmanned vehicles [3], and etc. One important challenge in this field is to increase energy transfer efficiency.

Avoiding system complexities and possible health risks of far-field radiative power transfer, quasi-static (low-frequency) electromagnetic field is more suitable for mid-range WET system [4]. Inductive coupling between two coils is one way to achieve energy transfer. However the energy transferred is limited in a short distance as the rapid decaying of magnetic flux with distance. In 2009, strongly coupled magnetic resonance (SCMR) scheme [5, 6] was proposed to increase coupling efficiency of two coils in a WET system.

Adding other devices in WET system is an another way. Magnetic super-lens with negative permeability was theoretically proven to be helpful in improving energy transfer efficiency [7]. This method has also been experimentally validated with the help of metamaterials [8-10].

Transformation optics (TO) serves as a theoretical tool for designing these devices. Since it was proposed in 2006 [11], TO has helped scientists controlling EM field in a pre-designed manner. From the perspective of TO, a super-lens can be designed by a space folding transformation [12]. Apart from super-lens, magnetic concentrators based on space compression transformation [13] was theoretically proposed to enhance energy transfer efficiency. With the help of metamaterials, such magnetic concentrator has been experimentally demonstrated [14].

In this paper, we apply the optical translational projector (OTP) based on the spatial translational transformation [15] to improve the energy transfer efficiency in WET systems. Such OTP can greatly enhance the energy transfer efficiency in WET systems (e.g. the enhancement factor can be more than 2 orders). We have designed a 3D reduced structure composed by layered isotropic magnetic materials to realize the proposed OTP in 10 *MHz*, whose performance is verified by 3D numerical simulations based on finite element method (FEM). We also numerically study the influence of the loss on the performance of the proposed OTP. Compared with previous methods, the proposed method in this paper can achieve a better enhancement of the energy transfer efficiency in WET systems.

## Results

Two copper circular antennas are utilized as a simplified WET system (see Fig. 1(a)). We set a unit external current density excitation along the tangential direction (i.e. $J_r=0$ $A/m^2$ and $J_\theta=1$ $A/m^2$) in the source coil. The other coil is used as a receiving coil. Energy is transferred from the source coil to the receiving coil via magnetic coupling. Fig. 1(b) shows that the designed OTP shell is added around the source coil to improve the energy transfer efficiency. The detailed design for the OTP and the realization method are given later.

The norm of the magnetic field produced by the source coil, which performs like that of a magnetic dipole, is shown in Fig. 2(a). When the source coil is covered by the OTP, the magnetic field is obviously altered (see Fig. 2(b)). A much stronger magnetic field is observed on right side of the coil, which means the OTP can help the source coil to transfer more energy to its right side where the receiving coil is located. This phenomenon can be explained from the perspective of TO later. The working frequency in all simulations is 10*MHz* in this study.

With the help of OTP, more magnetic flux is transferred to the receiving coil, which leads to a larger inductive current inside the coil, and a higher energy transfer efficiency. The surface current on the cross section of the receiving coil without OTP and with OTP is plotted in Figs. 3(a) and (b), respectively, which shows the inductive current on the receiving coil is greatly enhanced by the OTP. The inductive current is mainly on the edge of the coil due to skin effect.

Surface integration of current density on the cross section of the two coils are calculated. The ratio is plotted in Fig. 4, where *I* is the integration of current density on the receiving coil, and $I_s$ is the integration on the source coil. $I_s$ is a constant of $4.64 \times 10^{-7} A$. We can see that our OTP can greatly increase the current intensity inside the receiving coil. As distance between two coils increases, the inductive current on the receiving coil decreases. However the inductive current on the receiving coil is always enhanced by introducing our OTP (compared with the case without OTP).

To make a quantitative comparison with other methods, we define the enhancement factor of the energy transfer efficiency by:

$$A = \frac{P_{OTP}}{P_0}. \quad (1)$$

We use Joule heat produced on the receiving coil to measure the power harvested by the receiving coil [4]. $P_{OTP}$ and $P_0$ are Joule heat produced on the receiving coil with and without OTP, respectively. This definition consists with previous studies, in which power dissipated on the load is used to measure the power harvested by the receiving coil [4, 8, 14].

To realize the proposed OTP, we also make some simplification on the material requirement and propose a simplified OTP in which the negative permeability in the $z$ direction is set as 1. The numerical simulation results in Figs. 4 and 5 also show such a reduced OTP can also give a good coupling enhancement between two coils.

Metamaterials are required to realize the proposed OTP, as some regions of OTP (i.e. in the first and forth quadrants of our OTP) need negative permeability due to the spatial folding transformation (details will be explained later). The loss of resonate metamaterials to achieve a negative permeability will affect the performance of our OTP. We make some simulations on our OTP in a WET system when a small imaginary part $i\delta$ (i.e. loss) is added on the required permeability (see Fig. 6). Although the enhancement factor $A$ of the energy transfer efficiency (defined by Eq. (1)) decreases as the loss $\delta$ in the metamaterial increases, our OTP can still give a good energy transfer efficiency enhancement (i.e. $A>>1$).

## Method

To achieve a longer distance or a higher efficiency for a fixed distance in a WET system, we can use a device that can transfer more magnetic energy from the source coil to the receiving coil. In other words, this device should redirect the EM field generated by the source to a further distance. A lens usually can achieve such a function, since it can refocus the EM field of the source (e.g. a super-lens). Our aim is to design a special lens/shell that can project/shift the EM source (e.g. the source coil) by a pre-designed distance $d$ to another spatial position outside the lens where the receiving coil is located. It means that if an EM source is set in such a shell (i.e. our OTP), the EM field produced by the whole system outside the shell is the same as the field produced by an EM source located at the position shifted by a distance $d$ from its real position. The OTP shell shown in Fig. 7(a) can achieve such a function. The white and red regions are air due to the identity transformation (i.e. $x'=x$, $y'=y$, $z'=z$) and the spatial translational transformation along $x'$ direction (i.e. $x'=x-d$, $y'=y$, $z'=z$), respectively [15]. Here we use quantities with and without primes to indicate the quantities in the real and reference spaces, respectively, which consists with previous definition in TO [15]. The blue regions (i.e. our OTP) are filled with special media, in which the coordinate transformation is given by

$$x' = \begin{cases} -\dfrac{b-a}{d-b+a}x - \dfrac{d}{d-b+a}y + \dfrac{d}{d-b+a}b & \text{,for region in the first quadrant} \\ \dfrac{b-a}{d+b-a}x + \dfrac{d}{d-b+a}y - \dfrac{d}{d+b-a}b & \text{,for region in the second quadrant} \\ \dfrac{b-a}{d+b-a}x - \dfrac{d}{d+b-a}y - \dfrac{d}{d+b-a}b & \text{,for region in the third quadrant} \\ -\dfrac{b-a}{d-b+a}x + \dfrac{d}{d-b+a}y + \dfrac{d}{d-b+a}b & \text{,for region in the forth quadrant} \end{cases}, y' = y, z' = z, \quad (2)$$

where $a$ and $b$ are geometrical parameters of our OTP (see Fig. 1(b)). $d$ is the translational distance. With the help of TO, the required relative permeability in the blue region can be calculated [12]:

$$\mu' = \begin{bmatrix} \dfrac{(P^2+Q^2)}{P} & \dfrac{Q}{P} & 0 \\ \dfrac{Q}{P} & \dfrac{1}{P} & 0 \\ 0 & 0 & \dfrac{1}{P} \end{bmatrix}, \quad (3)$$

where $P=-sgn(x')\Delta/(d-sign(x')\Delta)$, $Q=-sign(x')sign(y')d/(d-sign(x')\Delta)$, and $\Delta = b-a$. Note that we have assumed the medium is air (i.e. the relative permeability is 1) in the reference space in the above calculation, and the permeability is uniform in each quadrant of OTP. As a low frequency EM source (e.g. $f$=10MHz) is adopted for the WET system, quasi-static approximation indicates that the electric field and magnetic field are decoupled. This is the reason why we only need to consider the relative permeability of our OTP (i.e. the relative permittivity is set as $\varepsilon_r$=1 in all simulations).

## Discussion

The function of the proposed OTP (described by Eq.(3)) can be understood from Fig. 7. From Figs. 7(b) to 7(f), the 2D spatial translational transformation carried out by the OTP (blue regions in figure 7(a)) from the real space to the reference space is given vividly: The square red domain $A_1B_1C_1D_1$, where the source coil is located, is stretched outside of the $x'$-$y'$ plane along $z'$ direction (see Figs. 7(b) to 7(c)). This region $A_1B_1C_1D_1$ is then shifted by a pre-designed distance $d$ along $x'$ direction from Figs. 7(c) to 7(d). Finally we press down $A_1B_1C_1D_1$ to the $x'$-$y'$ plane and obtain a spatially translated new square region $A_3B_3C_3D_3$ where the receiving coil is located. The function of the OTP is to shift the source coil by a pre-designed distance $d$ to a new spatial position (i.e. its image) where the receiving coil is located.

A super-lens with negative permeability (see Fig. 8) can also enhance the energy transfer efficiency in WET systems. It can also be designed by a spatial folding transformation (e.g. a 1D folding along $x$ direction) [12]. This leads to an infinitely large size of super-lens in both $y$ and $z$ directions (i.e. $B$ and $C$ in Fig. 8 should be infinitely long). A practical 3D super-lens with a finite length in both $y$ and $z$ direction (i.e., truncated in two directions) will greatly influence its performance. Our OTP utilizes 2D folding transformation, and thus it can achieve the full function in a 2D plane with a finite size. For a 3D OTP, it should be infinitely long only in $z$ direction (i.e. $h$ in Fig.1(b) should be infinitely long). Thus, for a 3D practical OTP we approximate only in $z$ direction (i.e., truncated only in one direction), which will also influence the performance of our OTP. However, for a 3D practical structure, our OTP makes truncation in only one direction (e.g. $z$ direction) while the super-lens makes truncation in two directions (e.g. both $y$ and $z$ directions). This is the reason why our OTP gives a much better performance than a super-lens, which will be shown in our numerical simulations later. To make a fair comparison between our OTP and a super-lens, we choose the same distance between two coils (e.g. $D_1=D_2=D/2=0.45m$ in Figs. 1(b) and 8). The theoretical image of the super-lens will be 0.9m away from the source coil, which coincides the image of our OTP since $d=0.9m$. To make a fair comparison, the thickness and height of the super-lens and OTP are chosen to be the same. The numerical simulations show our OTP can achieve a much higher efficiency than super-lens (see Fig. 9).

## Conclusion

In conclusion, an OTP designed by transformation optics is shown to be capable of greatly enhancing wireless energy transfer efficiency. Compared with a magnetic super-lens of the same size, our OTP can give a much better performance (e.g. the enhancement factors $A$ defined by Eq. (3) for the magnetic super-lens and our OTP are of about 2 times and 2 orders, respectively). 3D numerical simulations have been given to verify the performance of the proposed OTP. Even for the case where loss is considered, our OTP can still give a good performance. Our design will point out a new way for improving the energy transfer efficiency in WET systems.

## References

[1] Tesla, N. "Apparatus for transmitting electrical energy," *U.S. patent* number 1,119,732, issued in December (1914).
[2] Zhang, F. *et al.* Wireless energy transfer platform for medical sensors and implantable devices. *2009 Annual International Conference of the IEEE Engineering in Medicine and Biology Society* 1045–1048 (2009).
[3] Kim, J., Yang, S. Y., Song, K. D., Jones, S., Elliott, J. R., & Choi, S. H. Microwave power transmission using a flexible rectenna for microwave-powered aerial vehicles. *Smart Mater. Struct.* **15** (3), 889–892 (2006).
[4] Karalis, A., Joannopoulos, J. D., & Soljačić, M. Efficient wireless non-radiative mid-range energy transfer. *Ann. Phys.* **323** (1), 34-48 (2008).
[5] Kurs, A., Karalis, A., Moffatt, R., Joannopoulos, J. D., Fisher, P., & Soljačić, M. Wireless power transfer via strongly coupled magnetic resonances. *Science* **317** (5834), 83-86 (2007).
[6] Wei, X., Wang, Z., & Dai, H. A critical review of wireless power transfer via

strongly coupled magnetic resonances. *Energies* **7** (7), 4316-4341 (2014).

[7] Urzhumov, Y., & Smith, D. R. Metamaterial-enhanced coupling between magnetic dipoles for efficient wireless power transfer. *Phys. Rev. B* **83** (20), 205114 (2011).

[8] Lipworth, G., Ensworth, J., Seetharam, K., et al. Magnetic metamaterial superlens for increased range wireless power transfer. *Sci. Rep.* **4**, 3642 (2014).

[9] Ranaweera, A. L. A. K., Duong, T. P., & Lee, J. W. Experimental investigation of compact metamaterial for high efficiency mid-range wireless power transfer applications. *J. Appl. Phys.* **116**(4), 043914 (2014).

[10] Zhang, Y., Tang, H., Yao, C., Li, Y., & Xiao, S. Experiments on adjustable magnetic metamaterials applied in megahertz wireless power transmission. *AIP Adv.* **5** (1), 017142 (2015).

[11] Pendry, J. B., Schurig, D., & Smith, D. R. Controlling electromagnetic fields. *Science* **312** (5781), 1780-1782 (2006).

[12] Chen, H., Chan, C. T., & Sheng, P. Transformation optics and metamaterials. *Nat. mater.* **9** (5), 387-396 (2010).

[13] Navau, C., Prat-Camps, J., & Sanchez, A. Magnetic energy harvesting and concentration at a distance by transformation optics. *Phys. Rev. Lett.* **109**(26), 263903 (2012).

[14] Prat-Camps, J., Navau, C., & Sanchez, A. Quasistatic Metamaterials: Magnetic Coupling Enhancement by Effective Space Cancellation. *Adv. Mater.* **28**, 4898–4903 (2016).

[15] Sun, F., Liu, Y., & He, S. True dynamic imaging and image composition by the optical translational projector. *J. Optics* **18** (4), 044012 (2016).

[16] Sun, F. & He, S. Extending the scanning angle of a phased array antenna by using a null-space medium. *Sci. Rep.* **4**, 6832 (2014).

[17] Baena, J. D., Marqués, R., Medina, F., & Martel, J. Artificial magnetic metamaterial design by using spiral resonators. *Phys. Rev. B* **69** (1), 014402 (2004).

## Acknowledgment

This work is partially supported by the National Natural Science Foundation of China (Nos. 91233208and 60990322), the National Natural Science Foundation of China for Young Scholars (No. 11604292), the National High Technology Research and Development Program (863 Program) of China (No. 2012AA030402), the Program of Zhejiang Leading Team of Science and Technology Innovation, the Postdoctoral Science Foundation of China (No. 2013M541774), the Preferred Postdoctoral Research Project Funded by Zhejiang Province (No. BSH1301016), Swedish VR grant (# 621-2011-4620) and AOARD.

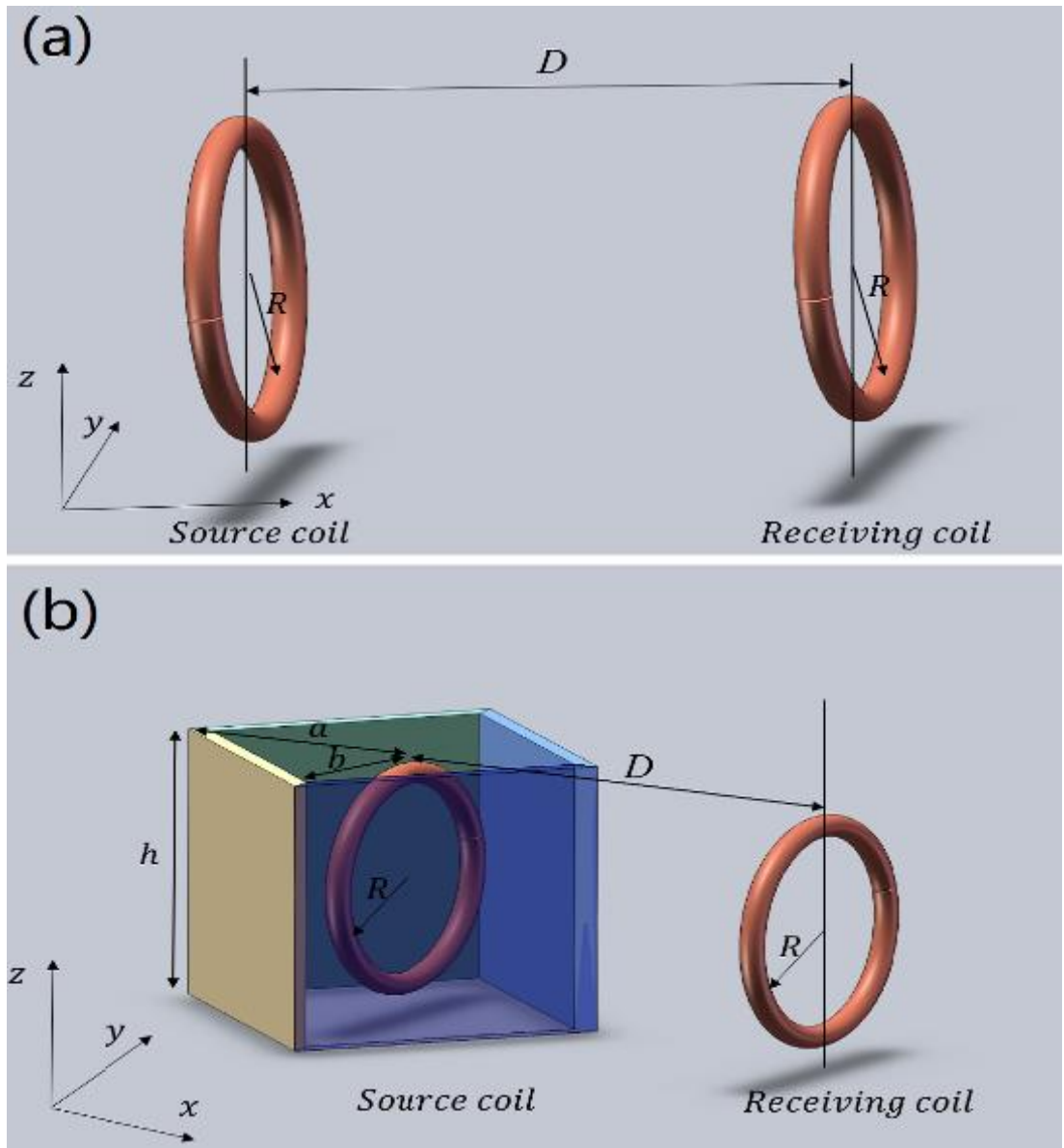

Figure 1| (a) Diagram of a simplified WET system. (b) Adding OTP in the WET system. *R* is the radius of the copper coil. *D* is the distance between the two coils. *a*, *b* are outer and inner size of the OTP, respectively. In this model, *a*=0.7*m*, *b*=0.8*m* *D*=1*m*,*h*=1*m* , and *R*=0.5*m*.

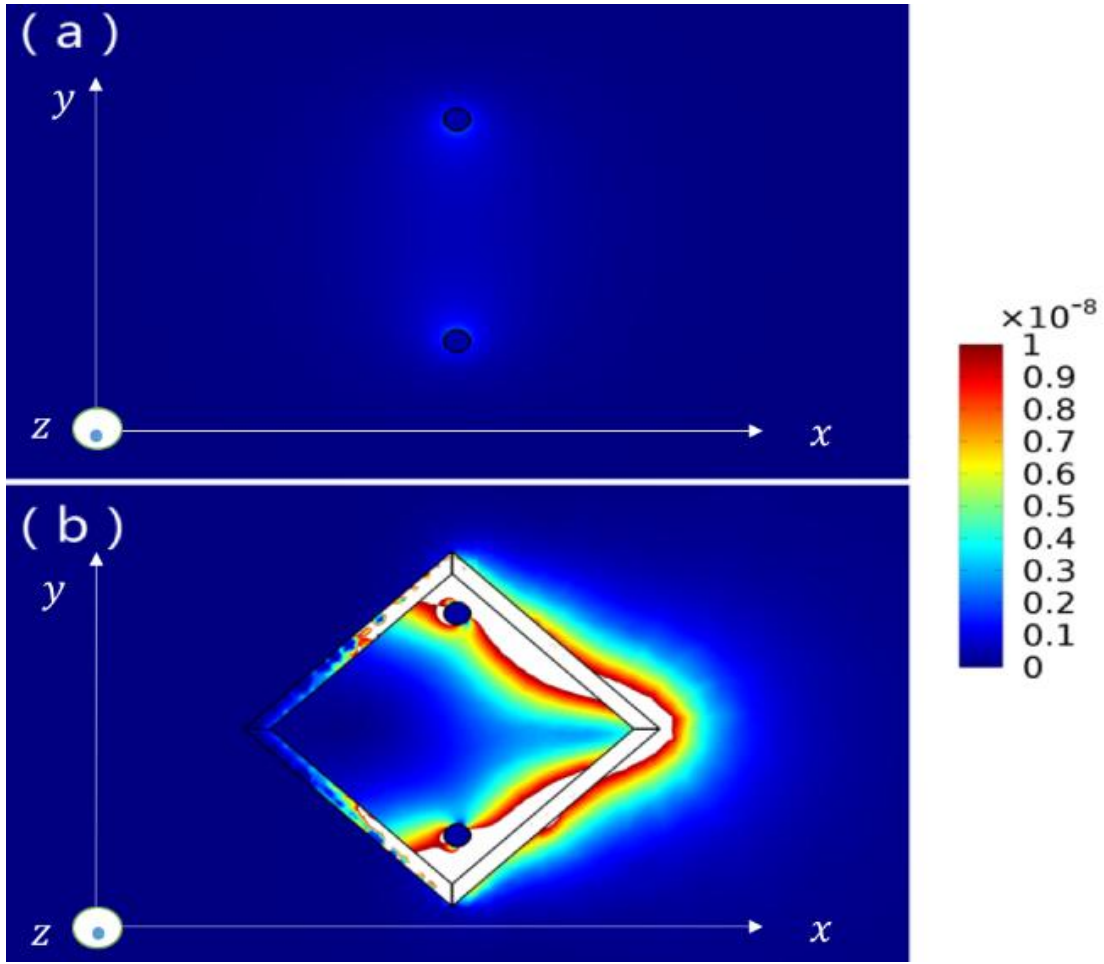

Figure 2| 2D simulation results. We plot the norm of magnetic field. (a) without OTP. (b) with OTP ($d=0.9m$, $a=0.7m$, $b=0.8m$, and $R=0.5m$.). Here white regions mean the field intensity are beyond the maxima of the color bar.

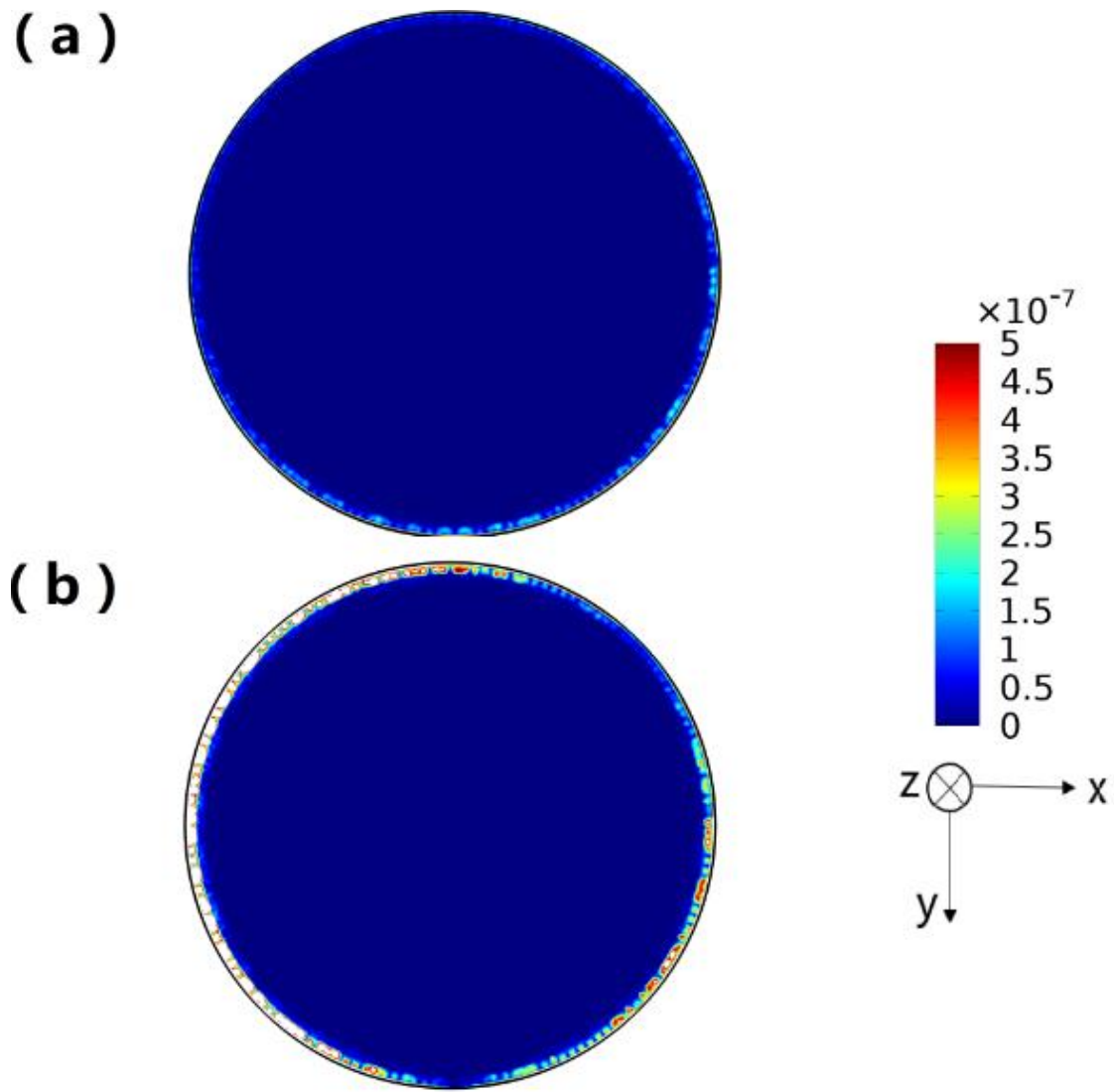

Figure 3| 2D simulation results. We plot the amplitude of surface current density on the cross section of the receiving coil. (a) without OTP. (b) with OTP. The geometrical parameters of OTP and two coils are the same as Fig. 2, and they remain unchanged in Fig. 4, 5, and 6. The white regions mean the field intensity are beyond the maxima of the color bar.

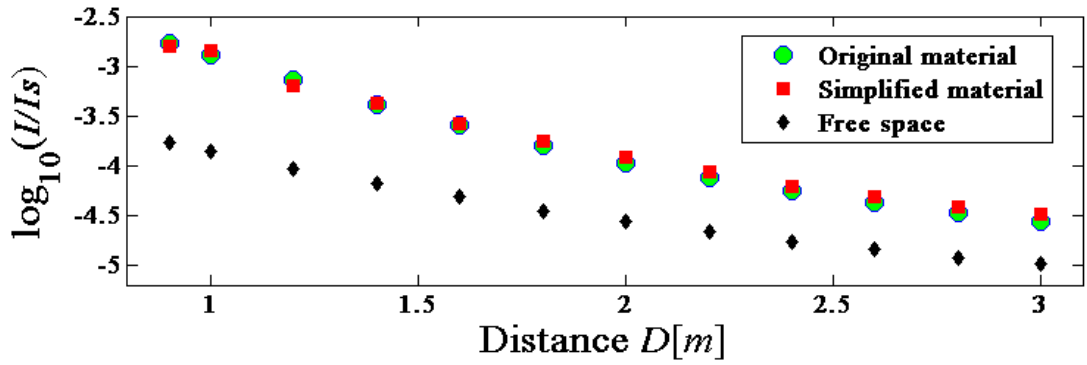

Figure 4| Ratio of surface current density integration as the distance $D$ between the two coils changes. $I$ is the current density integration on the cross section of the receiving coil, while $I_s$ is that of the source coil. The $z$-component of the permeability for the simplified material is reduced to 1.

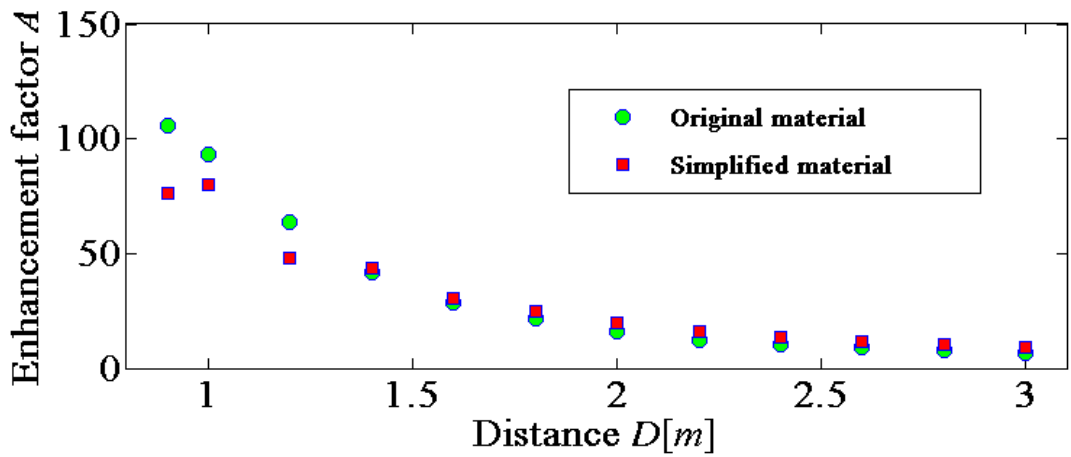

Figure 5| The enhancement factor $A$ of the energy transfer efficiency defined by Eq. (1) versus distance $D$ between two coils.

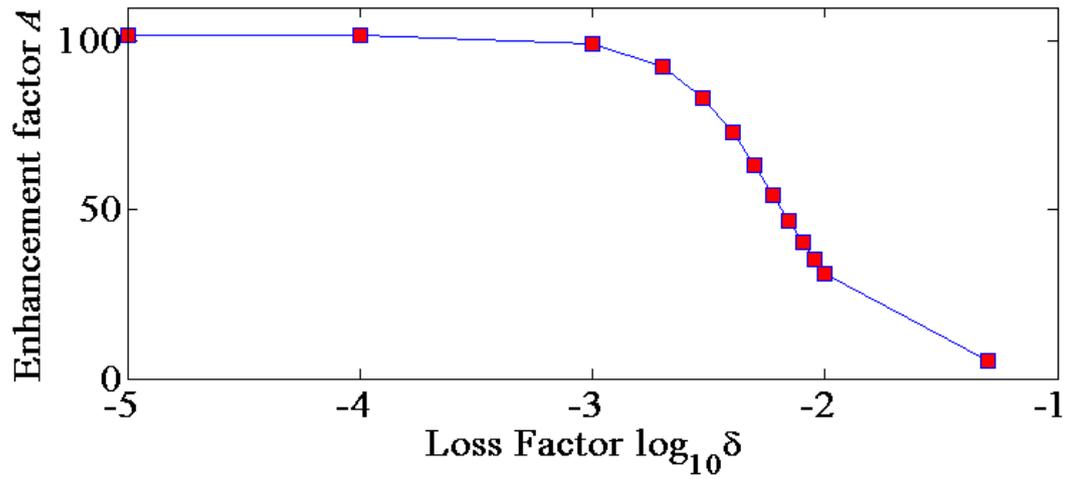

Figure 6| The relation between enhancement factor *A* and loss factor $log_{10}\delta$

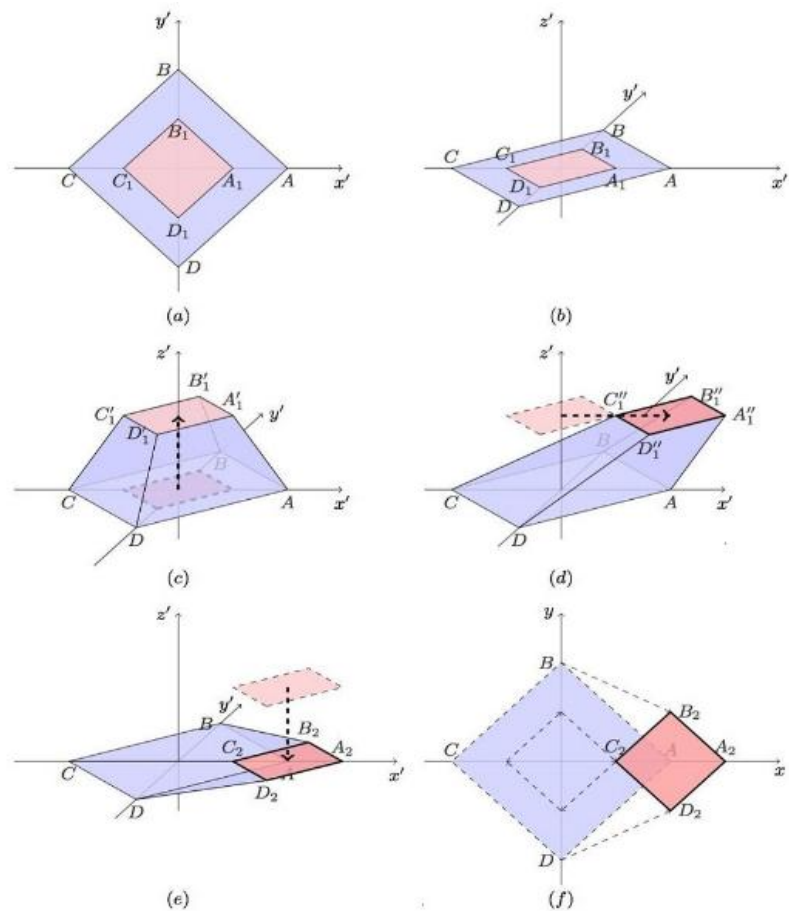

Figure 7| (a) The proposed OTP in the real space. (b)-(f) show the whole process of the spatial translational transformation from the real space to the reference space.

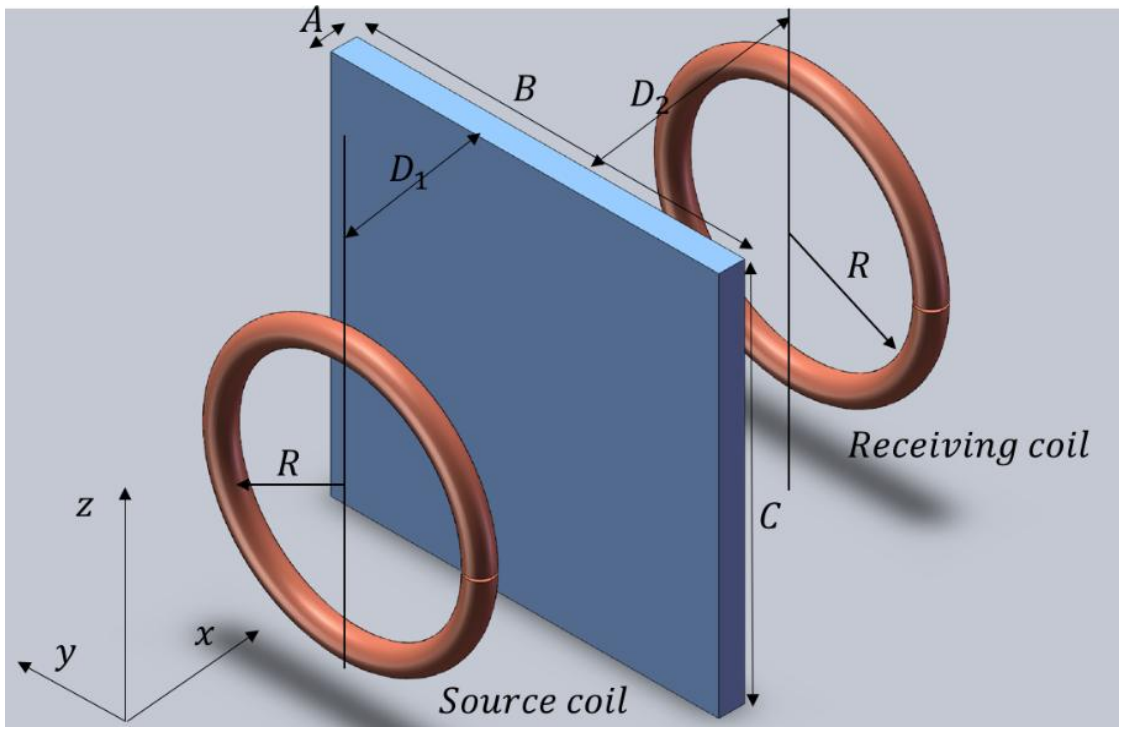

Figure 8| Diagram of a WTP system using a magnetic super-lens (i.e. the blue slab has $\mu_r = -1$).

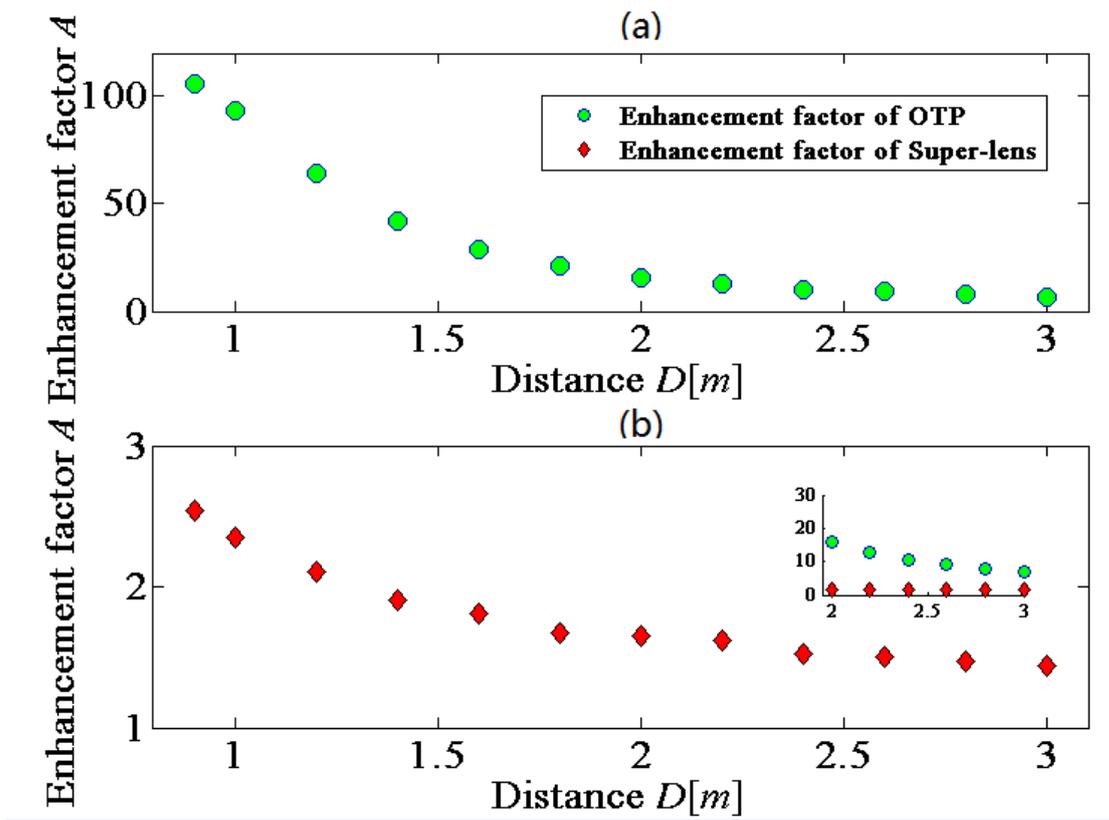

Figure 9| The enhancement factor $A$ in a WET system using (a) a 3D OTP with height

$h=1m$ (other parameters are the same as those for Fig. 2) and (b) a traditional super-lens with $A=0.1m$, and $B=C=1.5m$ in Fig. 8. The zoom-in inset shows the comparison when the distance $D$ exceeds $2m$.